\UseRawInputEncoding
\documentclass[prl,twocolumn,aps,showpacs,10pt]{revtex4-1}

\usepackage{graphicx}  
\graphicspath{{./Figures/}}
\usepackage{dcolumn}  
\usepackage{amssymb, amsmath}
\usepackage{natbib}
\usepackage{hyperref}

\newcommand{\PRLsection}[1]{\noindent \textit{#1} ---}

\usepackage{graphicx}
\usepackage{dcolumn}
\usepackage{bm}
\usepackage{color}

\definecolor{mygreen}{rgb}{0,0.5,0}
\definecolor{myred}{rgb}{0.75,0,0}
\definecolor{myblue}{rgb}{0,0,0.75}
\definecolor{mymagenta}{cmyk}{0,1,0,0.12}
\definecolor{mycyan}{cmyk}{1,0,0,0.12}
\definecolor{myorange}{rgb}{1,0.5,0}

\begin{document}

\title{Femtotesla nearly quantum-noise-limited pulsed gradiometer at Earth-scale fields}
\author{V.\ G.\ Lucivero}
\email[Corresponding author: ]{vito-giovanni.lucivero@icfo.eu}
\altaffiliation[Current affiliation: ]{ICFO-Institut de Ciencies Fotoniques, The Barcelona Institute of Science and Technology, 08860 Castelldefels (Barcelona), Spain}
\author{W.\ Lee}
\author{M.\ V.\ Romalis}
\affiliation{Department of Physics, Princeton University, Princeton, New Jersey, 08544, USA}
\author{M.\ E.\ Limes}
\author{E.\ L.\ Foley}
\author{T.\ W.\ Kornack}
\affiliation{Twinleaf LLC, Princeton, New Jersey, 08544, USA}
\date{\today}

\begin{abstract}
We describe a finite fields magnetic gradiometer using an intense pulsed laser to polarize a $^{87}$Rb atomic ensemble and a compact VCSEL probe laser to detect paramagnetic Faraday rotation in a single multipass cell. We report differential magnetic sensitivity of $14$ fT/Hz$^{1/2}$ over a broad dynamic range including Earth's field magnitude and common-mode rejection ratio higher than $10^4$. We also observe a nearly quantum-noise-limited behaviour of the gradiometer, by comparing the experimental standard deviation of the estimated frequency difference against the Cram\'{e}r-Rao lower bound in the presence of white photon shot-noise, atomic spin noise and atomic diffusion.
\end{abstract}


\maketitle

\PRLsection{Introduction}
Optically pumped magnetometers (OPMs), when operating at near-zero field in the spin-exchange
relaxation free (SERF) regime, are nowadays the most sensitive sensors to measure low-frequency magnetic fields reaching sub-femtotesla sensitivity \cite{Kominis2003,Dang2010} and surpassing that one of superconducting quantum interference devices (SQUID) \cite{David1972,Koch1993,Storm2016}. A major challenge in optical magnetometry \cite{Budker2007} is the extension of the SERF ultrahigh sensitivity to Earth-scale fields, where spin-exchange relaxation can only be partially suppressed \cite{Appelt1999}. This goal is important for a wide spectrum of applications, including space magnetometry \cite{DoughertyM.2006,Mateos2015,Korth2016}, magnetic navigation \cite{Shockley2014,Canciani2017}, archeological mapping \cite{David2004,Linford2007}, biomagnetism detection \cite{Bison2009,Jensen2016,Iivanainen2019}, mineral exploration \cite{Nabighian2005,Gavazzi2020}, searches for unexploded ordnance \cite{NelsonJune2001,Paoletti2019}, as well as for tests of fundamental physics \cite{Altarev2009,Lee2018}. In general, total-field OPMs can operate in geomagnetic fields (10-100 $\mu$T) \cite{Lee2021} and they are better suited for applications in challenging environments \cite{Fu2020}. Under continuous operation, finite field sensors based on Bell-Bloom (BB) \cite{Bell1961}, modulated nonlinear magneto-optical rotation (NMOR) and $M_z$ operation modes, have enabled sensitivity typically at the $\mathrm{pT/\sqrt{Hz}}$ \cite{Jimenez-Martinez2012,Lucivero2014,Patton2012} or slightly below that level \cite{Bevilacqua2016,Scholtes2011,Schultze2017}. A gradiometer with $1.5$ cm$^3$ active volume and sensitivity of 29 $\mathrm{fT/\sqrt{Hz}}$ at $26$ $\mu$T has been reported in \cite{Smullin2009}. At lower field of $7.3$ $\mu$T sub-femtotesla sensitivity has been demonstrated \cite{Sheng2013} using a gradiometer configuration in a single cm-sized multipass cell. A noise floor of about 100 $\mathrm{fT/\sqrt{Hz}}$ at $10$ $\mu$T has been obtained with a microfabricated 18 mm$^3$ cell \cite{Gerginov2020}. Recent works on intrinsic \cite{Lucivero2021,Perry2020,Zhang2020,Campbell2021} or synthetic \cite{Limes2020} (two spatially separated OPMs) gradiometers reported sensitivity below 50 $\mathrm{fT/cm/\sqrt{Hz}}$, enabling the first detection of human biomagnetism by OPMs in ambient environment \cite{Limes2020}.\\
Here we present a synchronous light-pulse atomic gradiometer \cite{Bell1961,Gerginov2017,Perry2020} operating in extreme conditions of pump power (few W per pulse) and pump pulses duration down to tens of nanoseconds. Using a single $0.5$ cm$^3$ multipass cell with adjacent compact probe and detection optics, we demonstrate differential magnetic sensitivity of 14 $\mathrm{fT/\sqrt{Hz}}$ over a broad range up to $50$ $\mu$T, Earth's field magnitude, and common-mode rejection ratio (CMRR) higher than $10^4$. We also report experimental saturation of the standard deviation of the estimated frequency difference versus fitting time, in good agreement with the Cram\`{e}r-Rao lower bound due to white photon noise plus additional atomic noise. The near quantum-limited sensitivity opens the possibility for quantum-enhancement in geomagnetic fields by squeezed-light \cite{Wolfgramm2010,Lucivero2016,Troullinou2021} and spin squeezing \cite{Appel2009,Wasilewski2010,Bao2020,Kong2020}.\\
\PRLsection{Sensor design and experimental setup}
\begin{figure}
	\centering
	\includegraphics[width=\columnwidth]{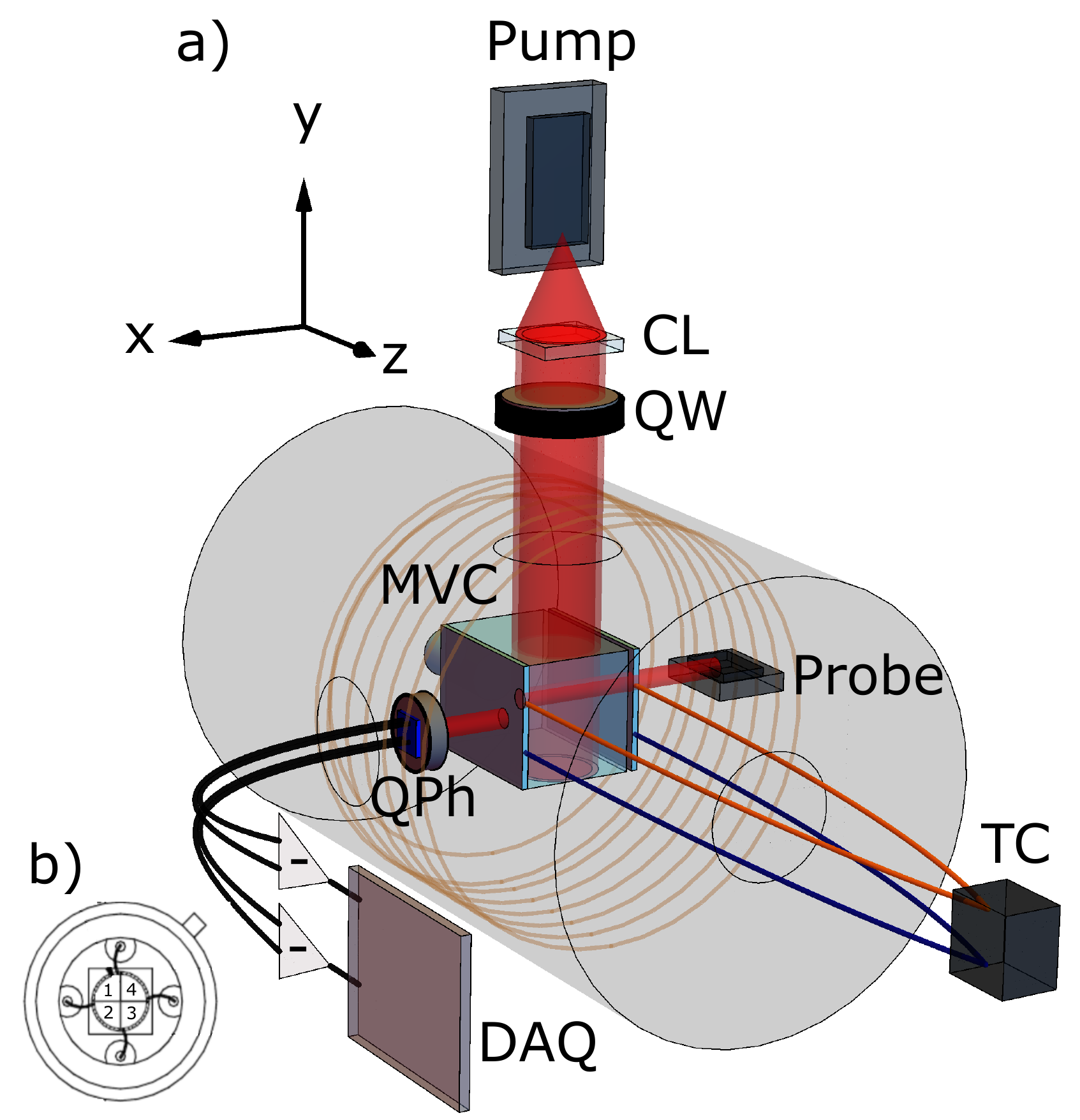}
	\caption{\textbf{a) Schematic of the experiment.} CL, Cylindrical Lens; QW, Quarter Wave-plate; MVC, Multipass Vapor Cell; TC, Temperature Controller; QPh, Quadrant photodiode; DAQ, Data acquisition card. \textbf{b) Photodiode Frontside view.} Active areas (1-4) of the quadrant photodiode.}
	\label{fig:setup}
\end{figure}
The full experimental apparatus is shown in Fig. (\ref{fig:setup}). The compact atomic sensor consists of a  multipass vapor cell, fabricated through anodic bonding of glass windows and internal mirrors \cite{NezihPatent}, fiberized optical heaters, a VCSEL probe laser, collimation and detection optics (not shown) and a quadrant photodiode. The cell is filled with pure $^{87}$Rb and $p_{\mathrm{N}_2}=700$ Torr of N$_2$ buffer gas pressure. The effective interaction volume is $62$ mm$^3$. The compact sensor is placed inside $5$ $\mu$-metal layers of magnetic shielding while a concentric set of cylindrical coils generate main field $B_z$ and first order gradient $\partial B_z/\partial y$. A semiconductor multi-mode pump laser is tuned to the $^{87}$Rb $D_1$ line, it is circularly polarized and aligned along the y-axis to maximize initial atomic polarization. The pump laser works in a pulsed regime with an averaged power of several W per pulse and adjustable repetition rate, number and width of pulses. When a finite $B_z$ field is applied, in order to produce a resonant build-up of atomic spin orientation \cite{Bell1961,Gerginov2017}, we synchronize the pump repetition time with the Larmor period $T_{pump}=1/\nu_L$, where $\nu_L=(\gamma/2\pi)B_z$ is the Larmor frequency and $\gamma=g_F\mu_B/\hbar$ is the gyromagnetic ratio. After a train of $N_{pump}=10$ pulses of $\tau_{pump}=1$ $\mu$s width, the free precession of atomic polarization is probed by a 100 GHz blue-detuned VCSEL laser which undergoes paramagnetic Faraday rotation and, after $15$ reflections across the multipass vapor cell, is split and detected by two balanced polarimeters consisting of compact polarization optics and four detectors, i.e. the active areas of a quadrant photodiode as shown in Fig. (\ref{fig:setup}-b). The pump-probe cycle occurs repeatedly at driving frequency of $f_d=180$ Hz.
\begin{figure}
	\centering
	\includegraphics[width=\columnwidth]{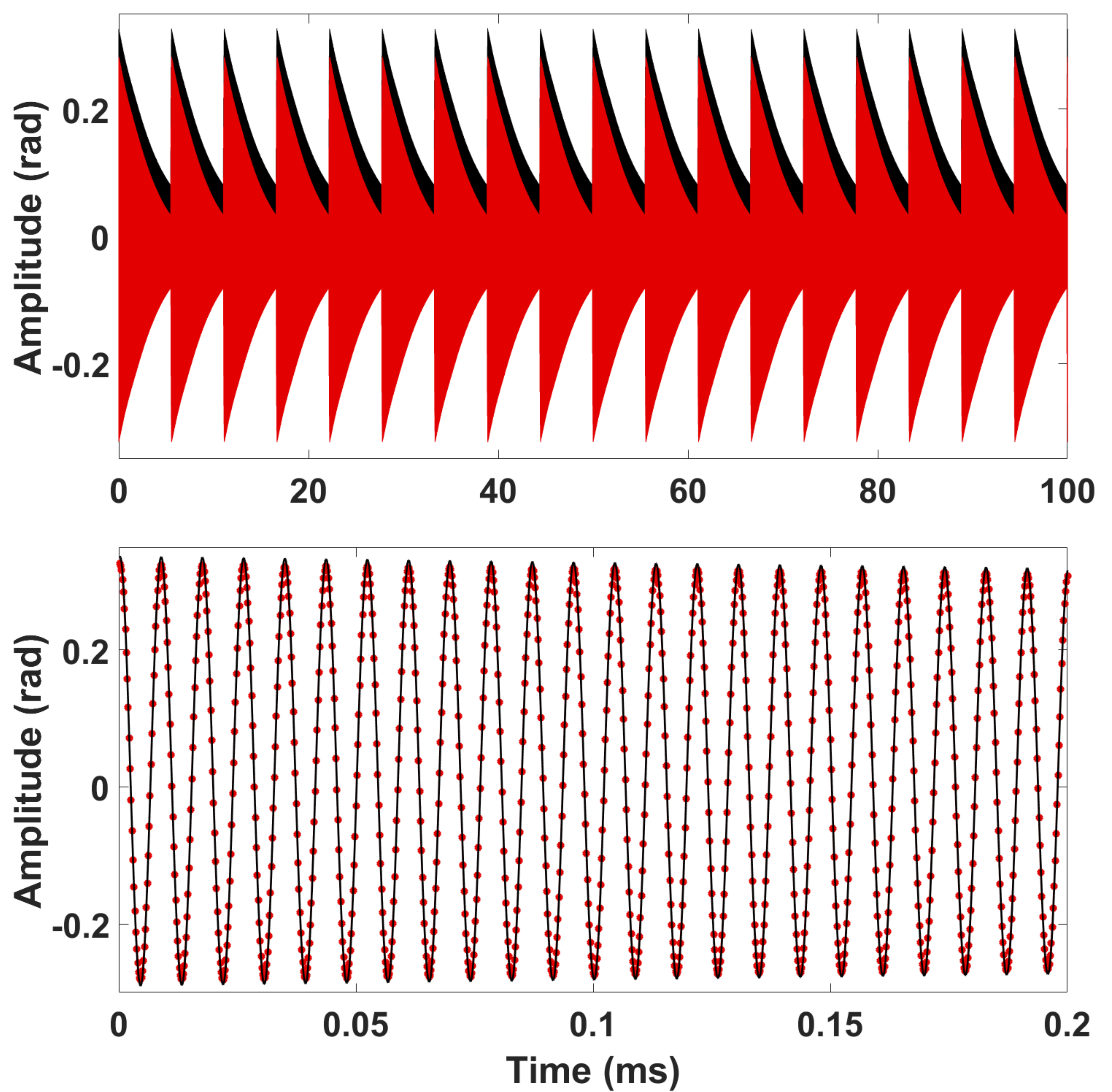}
	\caption{\textbf{(Top) Experimental signals.} Rotation signals of the two channels $V_{14}$ (back, black) and $V_{23}$ (front, red) acquired from the digital oscilloscope at field $B_z=16$ $\mu$T and $400$ $\mu$W total probe power. \textbf{(Bottom) Short time signal and fit} Zoom of the rotation signal (red points) and fit result (black continuous line) for the initial $0.2$ ms.}
	\label{fig:signals}
\end{figure}
After amplification with a transimpedance gain $G=50$k, the two differential signals $V_{14}=V_1-V_4$ and $V_{23}=V_2-V_3$ are fed into a 24-bit digital oscilloscope (DAQ). Such detection geometry enables a gradiometer operation mode with $0.2$ cm baseline, i.e. the spatial distance between top and bottom half of the quadrant photodiode. The mapping of atomic FID evolution onto light probe polarization, in absence of wrapping for small rotations ($\phi<\pi/4$) \cite{Li2011}, turn into detected signals with the form of a sine wave with exponential decay:
\begin{equation}
V_{14}(t)= V_{23}(t) = V_0 \sin(2\pi\nu_L t + \delta) e^{-t/T_2}
\label{signal}
\end{equation}
where $V_0$ is the maximum amplitude and $T_2$ is the transverse relaxation rate. In Fig. (\ref{fig:signals}) we report typical FID rotation signals from the two gradiometer channels and a sample fit to the function given in Eq. (\ref{signal}) for a shorter timescale. The temperature is optimized for sensitivity to $100^{\circ}$ C with a number density of $4.4\times10^{12}$ atoms/cm$^3$, measured via absorption spectroscopy \cite{Romalis1997}.\par
\PRLsection{Data analysis and results}
We acquired a FID train of signals for $T=20$ sec with $r_s=5$ MHz sample rate. Then, we fit all signals in each gradiometer channel for a duration $T_m=5$ ms to get a frequency difference array $\{\Delta\nu^{(i)}\}_{i=1,...,n}$. Each sample is temporally spaced by $\tau_d=1/f_d\approx 5.5$ msec.
\begin{figure}
	\centering
	\includegraphics[width=\columnwidth]{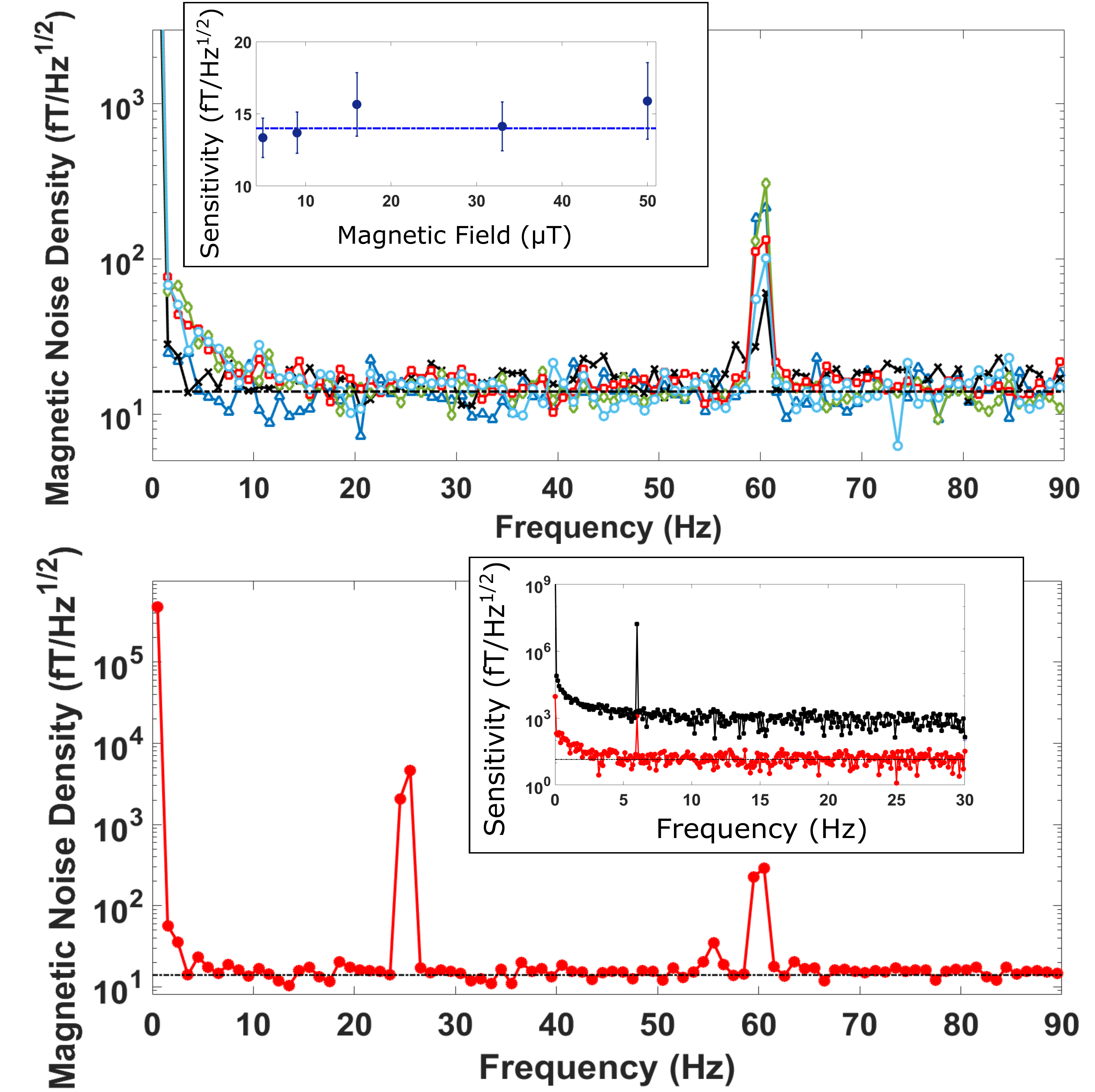}
	\caption{\textbf{(Top) Gradiometer sensitivity.} Experimental magnetic noise spectral density at finite fields of $5$ $\mu$T (blue triangles), $9$ $\mu$T (black crosses), $16$ $\mu$T (green diamonds), $33$ $\mu$T (red squares) and $50$ $\mu$T (cyan circles). \textbf{Inset.} Directly estimated (see text) magnetic noise density (blue points) versus main field magnitude. Sensitivity reference of 14 $\mathrm{fT/\sqrt{Hz}}$ (dot-dashed line) in both main plot and inset.
\textbf{(Bottom) Gradiometer signal.} Experimental noise spectral density for an applied $\partial B_z/\partial y$ gradient oscillating at $25$ Hz. \textbf{Inset: CMRR.} Magnetometer (black) and gradiometer (red) noise density for an applied $B_z$ oscillating at $6$ Hz.}
	\label{fig:gradsensCMRR}
\end{figure}
In Fig. \ref{fig:gradsensCMRR}(top) we report the results in the frequency domain, by showing the gradiometer magnetic spectral noise density $\rho_{\Delta B}(\nu)=(2\pi/\gamma)\sigma_{\Delta\nu}/\sqrt{\Delta f}$, i.e. the differential magnetic sensitivity in $\mathrm{fT/\sqrt{Hz}}$, at different finite fields, where $\sigma_{\Delta\nu}$ is the standard deviation of the estimated frequency difference and $\Delta f=1/(2T_m)$ is the gradiometer bandwidth. Apart from the technical noise peak at $60$ Hz we experimentally measured a sensitivity of 14 $\mathrm{fT/\sqrt{Hz}}$ over a broad dynamic range. In fact, the magnetic noise floor does not change when the main field $B_z$ is increased by an order of magnitude from $5$ $\mu$T up to $50$ $\mu$T, i.e. Earth's field magnitude. To keep the sensitivity high for fast precession, where $\nu_L\gg1/T_2$, and to avoid atomic depolarization by the pump, we gradually decrease the width $\tau_{pump}$ of pump pulses while increasing their total number, e.g. $N_{pump}=150$ and $\tau_{pump}=50$ ns at $B_z=50$ $\mu$T. The sensitivity can be alternatively obtained by directly computing the standard deviation of the frequency difference $\{\Delta\nu^{(i)}\}_{i=1,...,n}$ after averaging out the 60 Hz noise and calculating the short term difference between successive points. The results of this equivalent analysis are shown in the inset of Fig. \ref{fig:gradsensCMRR}(top) and agree within uncertainty with the 14 $\mathrm{fT/\sqrt{Hz}}$ spectral noise floor.\\
\begin{figure}
	\centering
	\includegraphics[width=\columnwidth]{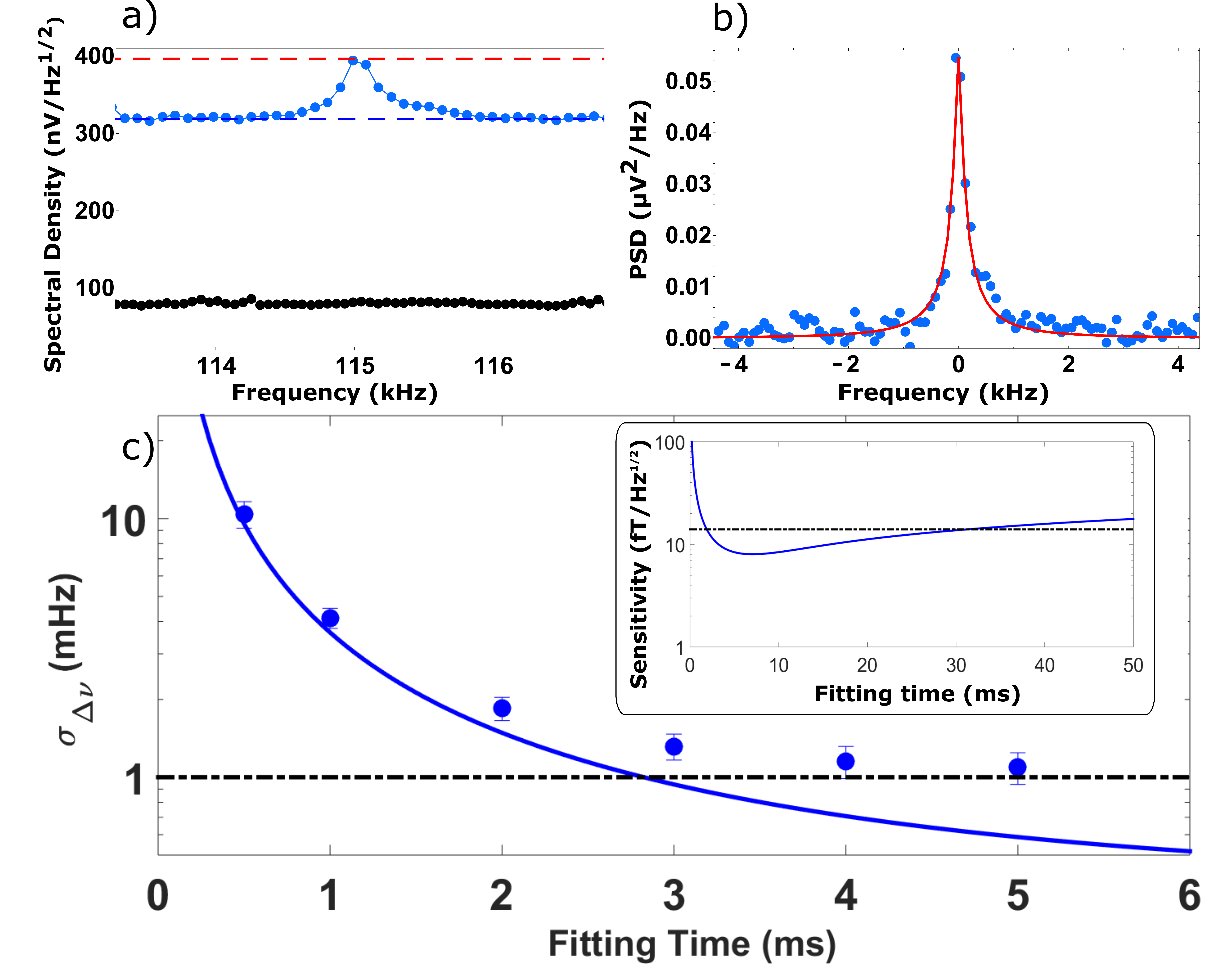}
	\caption{\textbf{a) Noise spectral density.} Experimental spin noise spectrum (blue points) at $B_z=16$ $\mu$T. Atomic noise peak amplitude (dashed red line) and photon shot-noise background (blue line) with probe beam on. Experimental electronic noise (black points) with probe beam off. \textbf{b) Normalized PSD.} Power spectral density (blue points) minus photon shot-noise background under same conditions of a). Analytical power spectrum (red line) for $700$ Torr of N$_2$ buffer gas and $0.2$ mm beam radius. \textbf{c) Uncertainty of frequency difference.} Experimental standard deviation of the gradiometer frequency difference (blue points) versus fitting time. Analytical CRLB (continuous blue curve) due to photon shot-noise and best experimental result (dot-dashed line). \textbf{Inset}: same curves of c) in magnetic units for longer fitting time.}
	\label{fig:Figure4}
\end{figure}
In Fig. \ref{fig:gradsensCMRR}(bottom) we report a sample signal of the atomic gradiometer, obtained by applying a $25$ Hz modulation to the magnetic coil gradient $dB_z/dy$ with amplitude of $0.8$ nT/cm. The figure shows the experimental noise spectral density with the applied gradient signal arising above the same noise floor of $14$ $\mathrm{fT/\sqrt{\text{Hz}}}$.
Furthermore, we measure the CMRR of the gradiometer by applying a $6$ Hz modulation to a dc magnetic field. The modulation amplitude is $1.4$ $\mu$T for the field parallel to the pump alignment $B_y$ and $1.8$ $\mu$T for the transverse field $B_z$. The ratio of the signal peak amplitude in the magnetometer (single channel) over the amplitude in the differential gradiometer spectral density gives the measured CMRR. In the inset of Fig. \ref{fig:gradsensCMRR}(bottom) we show the measurement when the modulation is applied in the $B_z$ main field. We measured a best CMRR of $1.2\times10^4$ and $2.9\times10^4$ for $6$ Hz modulation of the field in the two directions, respectively. This analysis shows good cancellation of broadband common field noise, as required for operation in unshielded ambient environment, which in fact has been demonstrated by using the same technique \cite{Lucivero2019QIM} in a portable sensor with two multipass cells separated by a 3 cm baseline \cite{Limes2020}.\par
\PRLsection{Fundamental theory and analysis} The gradiometer fundamental noise contributions are due to photon shot noise and atomic spin noise \cite{Budker2007}. In Fig. \ref{fig:Figure4}(a) we report the noise spectral density of a single gradiometer channel when the unpolarized ensemble is probed at $B_z=16$ $\mu$T. A typical spin noise spectrum \cite{Zap1981,Lucivero2017a} with peak noise density $\rho_{at}=3.9\times10^{-7}$ $\mathrm{V_{rms}/\sqrt{Hz}}$ arises above a white photon shot-noise background $\rho_{ph}=3.2\times10^{-7}$ $\mathrm{V_{rms}/\sqrt{Hz}}$, significantly above electronic noise level. The photon shot noise level agrees with its theoretical value \cite{Lucivero2014} $\rho_{ph}=\sqrt{2G^2e\Re P}$ where $e$ is the electron charge, $\Re=$ 0.57 A/W is the photodiode responsivity and $P=$ 200 $\mu$W is the total optical power incident on the single channel. In Fig. \ref{fig:Figure4}(b) we report the same noise spectrum in power density units, after subtracting white noise and shifting to zero frequency, and we compare it against the analytical power spectrum $S(\nu)=\langle V_{at}(t)^2\rangle\int_{-\infty}^{\infty}C_d(\tau)e^{-(\tau/T_2 + i 2\pi \nu \tau)}d\tau$, where $\langle V_{at}(t)^2\rangle=(\rho_{at}^2-\rho_{ph}^2)$ is the atomic noise power contribution and $C_d(\tau)$ is the contribution to the spin noise time-correlation function due to diffusion of atoms across the multipass probe beam \cite{Lucivero2017,Sheng2013}. We found good agreement against data with no free parameters and buffer gas pressure of $p_{N_2}=700$ Torr, beam width $w=0.2$ mm and physical $l_{cell}=10$ mm, where these quantities have been independently measured. This confirms that atomic diffusion broadens the Lorentzian spin noise spectrum giving a FWHM linewidth $\Delta\nu_c=1/(\pi T_c)$ of about 290 Hz, where $T_c$ is the noise correlation time including diffusion. This linewidth is about 3.2 times larger than that one due to the transverse relaxation time $T_2=3.5$ ms, obtained from the fit of the signals in Fig. (\ref{fig:signals}).\\
In order to study the quantum-noise-limited behaviour of the gradiometer sensitivity, in Fig. \ref{fig:Figure4}(c) we report the experimental standard deviation of the estimated frequency difference $\sigma_{\Delta\nu}$ versus fitting time. The theoretical Cram\`{e}r-Rao lower bound (CRLB) for the estimation of the frequency of an exponentially damped sinewave in the presence of white gaussian noise has been obtained in \cite{Yao1995,Gemmel2010}. The minimum bound for the variance is:
\begin{equation}
\sigma_{\nu,w}^2\geq \frac{12\rho_w^2C(T_m,T_2)}{(2\pi)^2A^2T_m^3}
\label{eq:CRBDampedSinewave}
\end{equation}
where $T_m=\Delta tN$ is the fitting time, $\Delta t$ is the discrete sampling time, $A$ is the signal amplitude and $\rho_w=N_w/\sqrt{f_{bw}}$ is the spectral noise density where $N_w^2$ is the gaussian noise variance and $f_{bw}=r_s/2$ is the detection bandwidth (or Nyquist frequency). The factor $C$, where $C=1$ for a pure sinewave, takes into account the exponential decay and, in the limit $\Delta t\rightarrow0$ (large $N$), is:
\begin{equation}
C(T_m,T_2)=\frac{2 T_m^3 e^{2\alpha} \left(e^{2\alpha}-1\right)}{3 T_2 \left(T_2^2 e^{4\alpha}+T_2^2-2 e^{2\alpha} \left(T_2^2+2 T_m^2\right)\right)}
\label{eq:Cfactor}
\end{equation}
where $\alpha=T_m/T_2$.
For a gradiometer measurement the final lower bound for the standard deviation of the estimated frequency difference is then given by:
 \begin{equation}
\sigma_{\Delta\nu,w}^{CRLB}\geq \sqrt{\sigma_{\nu1,w}^2 +\sigma_{\nu2,w}^2}
\label{eq:CRBTotal}
\end{equation}
where $\sigma_{\nu1,w}$ and $\sigma_{\nu2,w}$ are the lower bounds in the two channels in the presence of white noise only.
In Fig. \ref{fig:Figure4}(c) we compare the experimental standard deviation $\sigma_{\Delta\nu}$, measured up to $5$ ms, against the theoretical CRLB due to white photon shot noise. This is obtained with no free parameters by inserting in Eqs. (\ref{eq:CRBDampedSinewave}), (\ref{eq:Cfactor}) and (\ref{eq:CRBTotal}) experimental values of maximum signal amplitude $A=V_0$, spin relaxation $T_2=3.5$ ms and rms noise density $\rho_{ph}$ for both channels.
The discrepancy is due to the fact that Eq. (\ref{eq:CRBDampedSinewave}) does not include a second contribution due to the non-white atomic noise density $\rho_{at}$, which is comparable to the photon shot noise as the signal starts to decay above the spin coherence time $T_2=3.5$ ms. However, the same saturation of the experimental standard deviation occurs over the entire investigated dynamic range including $B_z=50$ $\mu$T. This demonstrates that the pulsed magnetic gradiometer operates in a nearly quantum-noise-limited sensitivity regime. Finally, in the inset of Fig. \ref{fig:Figure4}(c) we report the CRLB due to photon shot noise in magnetic sensitivity units, which shows an optimal region of fitting time due to the magnetometer's bandwidth scaling. However, the experimental sensitivity saturates for shorter times due to the atomic spin noise and the experimental fitting time is limited by the $180$ Hz driving frequency of the pump-probe cycle. The extension of Eq. (\ref{eq:CRBDampedSinewave}) to the presence of non-white noise is currently under theoretical investigation. \par
\PRLsection{Conclusions and outlook}
We demonstrated a compact magnetic gradiometer operating in pulsed mode at finite fields with $14$ fT/Hz$^{1/2}$ differential magnetic sensitivity over a broad dynamic range, $5$ $\mu$T to $50$ $\mu$T, up to Earth's field magnitude. This represents a significant step towards miniaturization of multipass-cell based \cite{Sheng2013} and highly-sensitive atomic sensors in geomagnetic fields. The high CMRR is suitable for applications in unshielded and challenging environments \cite{Limes2020,Fu2020}. The sensitivity could be further improved with higher VCSEL laser power and by using two pump lasers for hyperfine re-pumping \cite{Sheng2013,Schultze2015}, while a technique reducing heading errors due to the orientation of a total-field sensor has been recently developed \cite{Lee2021}. The operation mode of the described quantum-noise-limited gradiometer is compatible with quantum enhancement techniques using spin or polarization squeezing \cite{Shah2010,Troullinou2021} in geomagnetic fields. The analysis at fields above $50$ $\mu$T, due to non-linear Zeeman effect and hyperfine splitting, will require further investigation. \par
\PRLsection{Acknowledgements}
This work was supported by the Defense Advanced Research Projects Agency (DARPA). The views, opinions, and/or findings expressed are those of the authors and should not be interpreted as representing the official views or policies of the Department of Defense or the U.S. Government. During the manuscript writing, V. G. Lucivero has been supported by the H2020 Marie Sk{\l}odowska-Curie Action project PROBIST (Grant Agreement No. 754510).

\bibliographystyle{apsrev4-1}
\bibliography{FINFIELD}
\end{document}